\documentclass []{aa}
\usepackage {psfig,epsf,epsfig}
\newcommand{\uSiO}     {\mbox{$^{28}$SiO$\;$}}
\newcommand{\dSiO}     {\mbox{$^{29}$SiO$\;$}}
\newcommand{\tSiO}     {\mbox{$^{30}$SiO$\;$}}

\begin{document}

\def\th{$^{13}$}
\def\ei{$^{18}$}
\def\tw{$^{12}$}
\def\Lcs{{\hbox {$L_{\rm CS}$}}}
\def\lunits{K\thinspace \kms\thinspace pc$^2$}

\def\,{\thinspace}
\def\etal{et al.}

% units of measurement

\def\kms{km\thinspace s$^{-1}$}
\def\Lsun{L$_\odot$}
\def\Msun{M$_\odot$}
\def\ms{m\thinspace s$^{-1}$}
\def\percc{cm$^{-3}$}

% things with small caps  (watch out-- only for 12pt)
\font\sc=cmr10

\def\CBR{{\rm\sc CBR}}
\def\FWHM{{\rm\sc FWHM}}
\def\HI{{\hbox {H\,{\sc I}}}}
\def\HII{{\hbox {H\,{\sc II}}}}

% molecules and atoms and spectra

\def\Ha{H$\alpha$}                      % Hydrogen
\def\Htwo{{\hbox {H$_2$}}}
\def\nHtwo{n(\Htwo)}
\def\He#1{$^#1$He}                      % He isotopes
\def\water{H$_2$O}
\def\flecha{\rightarrow}
\def\12CO{$^{12}$CO}
\def\COJ#1#2{{\hbox {CO($J\!\!=\!#1\!\rightarrow\!#2$)}}}
\def\CO#1#2{{\hbox {CO($#1\!\rightarrow\!#2$)}}}
\def\CeiO#1#2{{\hbox {C$^{18}$O($#1\!\rightarrow\!#2$)}}}
\def\CSJ#1#2{{\hbox {CS($J\!\!=\!#1\!\rightarrow\!#2$)}}}
\def\CS#1#2{{\hbox {CS($#1\!\rightarrow\!#2$)}}}
\def\HCNJ#1#2{{\hbox {HCN($J\!\!=\!#1\!\rightarrow\!#2$)}}}
\def\HCN#1#2{{\hbox {HCN($#1\!\rightarrow\!#2$)}}}
\def\HNC#1#2{{\hbox {HNC($#1\!\rightarrow\!#2$)}}}
\def\HNCO#1#2{{\hbox {HNCO($#1\!\rightarrow\!#2$)}}}
\def\HCOpJ#1#2{{\hbox {HCO$^+$($J\!\!=\!#1\!\rightarrow\!#2$)}}}
\def\HCOp#1#2{{\hbox {HCO$^+$($#1\!\rightarrow\!#2$)}}}
\def\HCOpp{{\hbox {HCO$^+$}}}
\def\J#1#2{{\hbox {$J\!\!=\!#1\rightarrow\!#2$}}}
\def\noJ#1#2{{\hbox {$#1\!\rightarrow\!#2$}}}

% luminosities, etc

\def\Lco{{\hbox {$L_{\rm CO}$}}}
\def\Lhcn{{\hbox {$L_{\rm HCN}$}}}
\def\Lfir{{\hbox {$L_{\rm FIR}$}}}
\def\Ico{{\hbox {$I_{\rm CO}$}}}
\def\Sco{{\hbox {$S_{\rm CO}$}}}
\def\Ihcn{{\hbox {$I_{\rm HCN}$}}}

% radio astronomy

\def\Tastar{{\hbox {$T^*_a$}}}
\def\Tmb{{\hbox {$T_{\rm mb}$}}}
\def\Tb{{\hbox {$T_{\rm b}$}}}

\title{\uSiO, \dSiO and \tSiO excitation: effects of infrared line overlaps}

\author {
F. Herpin\inst{1}$^{,}$\inst{2},
A. Baudry\inst{1}
}

\offprints{F. Herpin, herpin@isis.iem.csic.es}

\institute {$^{1}$Observatoire de Bordeaux, BP 89, F-33270 Floirac, France \\
  $^{2}$Present address: Dept F\'{\i}sica Molecular, I.E.M., C.S.I.C, 
Serrano 121, 28006 Madrid, Spain}

\thesaurus{07 (02.12.1, 02.13.3, 08.03.4, 08.12.1)}

\date{Received ; accepted }

\maketitle
\markboth{Effects of infrared line overlaps between SiO isotopes}
{Effects of infrared line overlaps between SiO isotopes}
\begin{abstract}

We have investigated the influence of several 
line overlaps between \uSiO, \dSiO and \tSiO on maser emission 
from the envelopes of late-type stars using the 
{\em Large Velocity Gradient} radiative transfer formalism modified 
to account for overlaps. Our model includes: (i) 40 rotational levels 
for each of the $v=0$ to 4 vibrational states and each isotopic species; 
(ii) collisions with molecular and atomic hydrogen; and (iii) radiative  pumping 
by a shell of dust grains surrounding the SiO masing layers.

We have shown that several line overlaps may play a role on 
the excitation of the \uSiO $v=3, 
J=1\rightarrow 0$ maser, and on the anomalously weak $v=3, J=2\rightarrow 1$ 
emission. We suggest that \uSiO maser emission in the $v=4, J=5\rightarrow 4$ 
line, the only one observed in this vibrational level, may 
also result from line overlaps. 
For \dSiO, infrared line overlaps successfully explain observed maser emission in the 
$v=0, J=5\rightarrow 4$ and $v=1, 2, 
J=6\rightarrow 5$ lines. 
\tSiO, $v=0, J=2\rightarrow 1$, and $v=0, J=4\rightarrow 3$ 
maser emissions could also result from line 
overlaps, and we find that several overlaps could explain the newly 
discovered maser line \tSiO $v=1, J=1\rightarrow 0$. Finally, we make some predictions for new \uSiO, \dSiO and 
\tSiO maser lines.


\end {abstract}

\section{Introduction}

SiO masers are excited in the inner layers of circumstellar 
envelopes of late-type stars, before the dust formation zone, but 
beyond the photosphere where outflow and inflow of matter or shocks 
coexist and give rise to complex physical conditions. 

Since the discovery of SiO maser emission by Snyder \& Buhl
(1974), many papers have been published to explain the SiO maser
phenomenon, and several improvements involving collisional and
radiative pumping have been proposed to model
 the emission from late-type stars. However, it is still difficult to
explain the disparities between some rotational line 
intensities observed inside a same vibrational
state or, more generally, in adjacent transitions
of \tSiO, \dSiO and $v \geq 3$ \uSiO.
Olofsson \etal (1981, 1985) introduced the hypothesis
of a line overlap
between two ro-vibrational lines of SiO and water to
explain the weakness of the \uSiO $v=2, J=2\rightarrow 1$
emission. This emission was observed by Bujarrabal \etal (1996)
in stars with different O-abundances, and the overlap between two infrared lines
of H$_{2}$O and SiO was confirmed. 
In addition, several overlaps between the infrared lines of the main 
(\uSiO) and rare (\dSiO and \tSiO) isotopic species of silicon 
monoxyde have been suggested to play a role in the excitation of 
these species. Using the 
{\em LVG} approximation Cernicharo \etal (1991) modeled the pumping 
of \dSiO by \uSiO. The importance of overlaps was further 
investigated by Cernicharo \& Bujarrabal (1992) and 
Cernicharo \etal (1993). 
Recently, Gonz\'{a}lez-Alfonso \etal (1996) and Gonz\'{a}lez-Alfonso 
\& Cernicharo (1997) used a non-local radiative transfer code 
to study the question of line 
overlaps. They successfully explained  most 
of the \dSiO, \tSiO and high-v ($v\geq 3$) \uSiO masers; however, several 
lines remain unexplained. 

Our main goal in this work is to develop a simple and modular calculation tool 
in order to investigate the impact of various line overlap 
effects among the three main isotopic forms of SiO. We do not 
intend here to realistically model the SiO maser line profiles. We rather are 
interested in predicting and discussing relative peak line intensity ratios. 
In Sect. 2 we give details on 
spectroscopy, the used radiative transfer code and 
physical parameters, and on validation of our code. In Sect. 3, we 
include several line overlaps and compare our 
results to the observations. Some major results of this work are summarized in 
Sect. 4. 

\section{The Model}

The LVG code used here must be regarded as a simple but powerful
computational tool for maser emission prediction.
Bujarrabal (1994) has clearly demonstrated that the LVG code
applied to the SiO maser zone is a fairly good
approximation in comparison with the exact radiative
transfer solution. Other uncertainties do not play a minor role in 
comparison with the differences between the LVG and the exact radiative 
transfer results; in particular uncertainties in the collision laws and 
the exact geometry of the maser region result in weaker predictions. 
Thus, we believe that the physical laws and the spherical symmetry adopted 
in our model are sufficient for the LVG approximation.

\subsection{Spectroscopy, collisions and radiative transfer}
\label{spectro}

\hspace {0.5 cm}Modeling the SiO emission requires the 
use of accurate spectroscopic data and collisional rates. 
Energy levels of SiO have been calculated using the most recent 
calculations of the Dunham coefficients (Campbell \etal ~1995). 
For the Einstein A-coefficients we used the recent work of 
Drira \etal (1997).
                                   
Basic data concerning the rotational and rovibrational 
rate coefficients of SiO with H$_{2}$ and their dependence on temperature
were taken from Bieniek \& Green (1983). We also used the new rate
coefficients derived by Lockett \& Elitzur (1992)
for $v\geq 3$ and $\Delta v=2$ transitions and we
applied corrections to their rates
to compensate for the effect of the limited number of rotational and 
vibrational levels taken in the calculations. In addition,
Langer \& Watson (1984) made a first attempt to estimate collision rates 
of SiO with atomic hydrogen from a comparison with the expected 
H-CO and H$_{2}$-CO rates. The main features of these H-SiO 
rates used in our work are: the vibrational collision rates are 
10 times larger than those for H$_{2}$, and 
the rotational collision rates are decreased by a factor of 7. 
H can be formed in two 
different ways: ({\em i}) the stellar radiation may photodissociate H$_{2}$ close to the star, 
thus forming an atomic hydrogen layer; ({\em ii}) shocks may dissociate 
$H_{2}$ to produce a uniform distribution of H throughout the envelope. 
We assume here that $\chi [H_{I}/H_{2}]$ is constant and equal to 
$10^{-2}$ as in Bowers \& Knapp (1987). 
Adding atomic hydrogen to H$_{2}$ increases the role of collisions. 
Note however that its impact on the
inversions is weak (a few \% for $\chi [H/H_{2}]
<10^{-1}$). Beyond this value, inversions for the lowest rotational 
levels tend to decrease 
contrary to what is observed for high $J$ transitions. The higher the 
vibrational level, the larger the effect of the atomic 
hydrogen is; this is expected because of larger vibrational collision rates 
for H than for H$_{2}$. Nevertheless, we stress that these collision 
rates and the relative abundance 
of the atomic hydrogen are poorly known.
\label{transclas}

Because the purpose of this work is to explain in a simple
way the main characteristics of \uSiO, \dSiO and \tSiO emissions,
we have used
the {\em LVG} (Sobolev 1958) approximation, based on the model of a
homologously expanding envelope, 
in which the escape probability does not depend on the angle.
If the velocity gradient (logarithmic velocity gradient 
$\varepsilon_{r}= d\ln V/d\ln r$) 
in the circumstellar envelope is large enough, each cell of 
the discretized medium will 
be decoupled from all other cells, thus allowing a local 
treatment of the radiative transfer. The interaction area is 
limited to a small zone where photons can be absorbed or emitted 
around a resonance point. 
We have limited the maximum 
size of the cells to 
the {\em Sobolev length}, $L$, defined by 
\[L=r \frac{\Delta V_{D}/V}{\varepsilon_{r}}\]
with \[\Delta V_{D}=\sqrt{{\Delta V_{th}}^{2}
+{\Delta V_{turb}}^{2}}\] 
where $r$ is the radial distance, $\Delta V_{th}$ the thermal velocity,
  $\Delta V_{turb}$ the turbulence, and $V$ the expansion velocity.
$\Delta V_{D}$ characterizes the local absorption coefficient. The formula
giving the Sobolev length applies to any form of expansion velocity and to
the general case in which the photon escape probability depends on the
direction. For a spherically symmetric envelope and for the special
case where $V$ is proportional to $r$, the angle-dependent terms vanish,
the logarithmic gradient $\varepsilon_{r} =1$, and the escape probability is 
isotropic and proportional to the inverse of the peak opacity (as soon as the 
opacity becomes large). The Sobolev length is just the length over 
which the velocity varies by an amount corresponding to the width 
of the local absorption coefficient. 
One essential advantage of the LVG code is that it requires less computer time than more 
exact treatments of the radiative transfer problem based on integral 
solutions (Bujarrabal 1994) or on the Monte Carlo method (Gonz\'alez-Alfonso \& Cernicharo 
1997). In the frame of the LVG formalism and
for physical conditions relevant to the circumstellar
environment of late-type stars we have derived the solutions
of the statistical equilibrium equations and the opacities of
various SiO lines. More complex effects related to non 
spherical geometry, polarization or saturation and beaming angles 
are beyond the scope 
of this paper. Concerning saturation, we emphasize that our rate equations are 
solved for both positive and negative opacities, but that beaming related 
to saturation cannot be investigated here because we assume the escape 
probability to be isotropic. Note that beaming angles for saturated masers in the case of 
the {\em LVG} approximation remain an unsolved problem (Elitzur 1992). 
Saturation and the effect of competitive gains in SiO lines 
were investigated in Doel \etal (1995). 

Our calculations include 40 
rotational levels for each of the 5 vibrational states $v=0$ to 4. 
To correct for the limited number of 
rotational and vibrational levels (this overestimates the opacities), 
we have applied a correction similar to that used by Bujarrabal \& 
Rieu (1981) for the rotational populations. 
The required final precision on the populations is better than $10^{-4}$. 
Calculations are done simultaneously for \uSiO, \dSiO and \tSiO. 
We derive the populations for each level ($v,J$) and derive the opacities 
for all allowed transitions. 

\subsection{Circumstellar parameters}

We have used physical parameters appropriate to the circumstellar 
environment of evolved stars (see details in Herpin 1998).

\label{paraphy}
\paragraph{$*$ Kinetic temperature \\}

The gas temperature follows the law adopted by Langer \&
Watson (1984): 
\[T_{c}=T_{\star} ({\frac{r}{R_{\star}}})^{-0.5}\]
where $R_{\star}$ is the stellar
radius ($R_{\star}=7.7~10^{13}$cm), $T_{\star}$ is the stellar
temperature ($T_{\star}=2500$ K) and r is the radial distance. 
This law is in rough agreement with the observations of SiO 
thermal emission of Lucas \etal (1992) and Bujarrabal \etal (1986), 
and also with the calculations of Willson (1987) 
who finds 1500 K in the maser region, for a typical Mira. 
Turbulence is a free parameter in our 
code and contributes to net line broadening. 

\paragraph{$*$ Expansion velocity \\}

\label{parvit}

The expansion velocity of the gas adopted in our model is: 
\[V=V_{\infty}\frac{1}{1+(\frac{R_{acc}}{r})^{e_{v}}}\]
where $R_{acc}=1.8~10^{15}$ cm, $e_{v}\in [1;3]$ and $V_{\infty}$ is 
the asymptotic expansion velocity of the gas (9.5 \kms). 
A logarrithmic velocity gradient of 3 was used in the calculations.

\paragraph{$*$ SiO abundance \\}

%
%Figure
%
\begin{figure} [ht]
  \begin{center}
     \epsfig{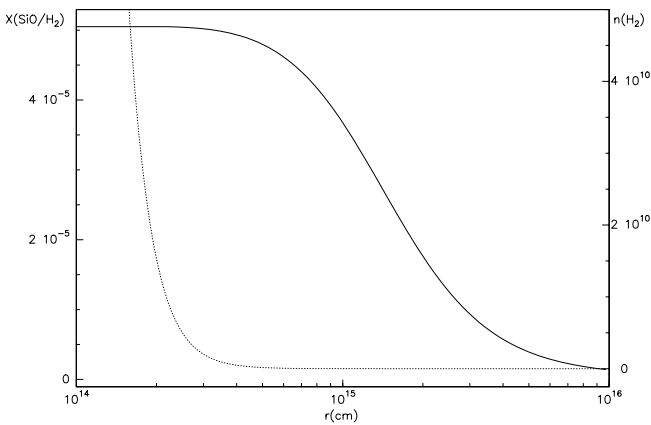}
   \end{center}
  \caption []{SiO$/$H$_{2}$ abundance $\chi$ (continuous line) and
    H$_{2}$ density (in cm$^{-3}$, dotted line) versus the
distance to the star $r$(cm) for $\chi_{0}=5~10^{-5}$,
$\frac{dM}{dt}=5.3~10^{-7}$ M$_{\odot}/$yr and $R_{\star}=7.7~10^{13}$ cm.}
  \label{chi}
\end{figure}

The SiO$/$H$_{2}$ abundance ratio is uncertain, and we adopt the solar
abundance by default ($\chi_{0} [$SiO$/$H$_{2}]= 5.~10^{-5}$). 
This abundance decreases with the distance to the star, the SiO molecules
being progressively condensed onto the grains. 
We follow the law adopted by Bujarrabal \etal (1989),
and we adopt for the \dSiO and \tSiO isotopic abundances ($A_{i}$) 
relative to \uSiO, $1/19.5$  and $1/29.5$, respectively (see fig.\ref{chi}): 
\[\chi (r)=\chi_{0}[1-\frac{\kappa (r)}{\kappa_{0}}+\chi']/A_{i}\]
with $\chi_{0}$ the initial ratio (here the solar abundance), and
$\chi'$ the remaining fraction of SiO after the grain formation ($10^{-2}$).
The terms $\kappa (r)$ and $\kappa_{0}$ depend respectively on the radiative 
pressure and the total quantity of grains formed.
We take: 
\[\kappa (r)=\kappa_{0} \frac{(r-R_{1})^{2}}{(r-R_{1})^{2}+(R_{d}-R_{1})^{2}}\]
where $R_{d}$ is the radius where half the total quantity of
grains is formed ($1.5~10^{15}$ cm),
and $R_{1}$ is the initial point ($2.~10^{14}$ cm).

\paragraph{$*$ \Htwo density \\}

We adopt (Elitzur 1992): 
\[n_{H_{2}}=\frac{\dot{M}}{8\pi r^{2}Vm_{p}}\]
with $V$ the expansion velocity, $\dot{M}$ the steallar mass loss rate
and $m_{p}$ the proton mass (see Fig.\ref{chi}).
Maser effects will require densities of a few $10^{9}$ particles/cm$^{3}$,
whereas thermalization is clearly reached beyond $10^{10}-10^{11}$
particles/cm$^{3}$.

\paragraph{$*$ Grain radiation \\}
\label{contgr}

IR spectra from the envelopes of late-type stars show that circumstellar 
grains have various chemical 
compositions. Here we only consider silicate grains and 
we assume that most circumstellar grains lie in a shell outside the SiO
maser zone (with $R_{in}=10^{15}$ cm, and $R_{out}=10^{16}$ cm). 
The extinction coefficient of the dust is 
$Q_{\nu} \simeq Q_{0}(\frac{\lambda}{\lambda_{0}})^{-p}$ 
with $p\in[1;2]$ for wavelengths $\lambda \geq 1 \mu$m (Mathis 1990). 
For silicate grains, we take $p=1.1$, $\lambda_{0}=80\mu$m, and 
$Q_{0}=2.10^{-3}$ according 
to the work of Ivezic \& Elitzur (1995). We consider 
that the silicate grains behave as black-bodies with temperature $T_{d}$. 
We assume that the dust 
is optically thin, and that $T_{d}$ is constant throughout the 
envelope ($T_{d}=$600 K). 
We take, following Netzer \& Elitzur (1993), 
${\chi}_{d}=\frac{M_{dust}}{M_{H_{2}}}=10^{-2}$, $3.0~$g.cm$^{-3}$ 
for the grain volumic density, and $5.10^{-6}$cm for the grain radius.

\subsection{Line Overlaps}
\label{lesrecouvrements}

In this work, we only deal with local line overlaps due to thermal line 
broadening and local turbulence in successive circumstellar layers. 
For simplicity, two nearby spectral lines are
considered to overlap when $\Delta V \leq \Delta V_{0}$
\label{eq:recouvrement}
where $\Delta V$ is the difference in velocity between the line 
centers, and $\Delta V_{0}$ is an {\em a priori} fixed value.
In the examples discussed in Sect. 3 we have adopted $\Delta V_{0}=5$
\kms corresponding to turbulent inner layers
($\Delta V_{thermal} \simeq 1.2$ \kms).
Line overlaps are treated in a very simple manner.
For two nearby lines ${\nu}_{12}$ and ${\nu}_{34}$, 
we derive the opacities $\tau_{12}$, $\tau_{34}$, and the individual source 
functions $S_{12}$, $S_{34}$. 
Adding the absorption and emissivity coefficients of each line, 
the source function becomes
\[{S_{12}}'=S_{12}(\frac{\tau_{12}}{\tau_{12}+\tau_{34}})
+S_{34}(\frac{\tau_{34}}{\tau_{12}+\tau_{34}})\]
and the total opacity is now ${\tau_{12}}'=\tau_{12}+\tau_{34}$.
The average intensity is then given by:

\[{\overline{J}}_{12}=[1-{\beta_{12}}'(\tau ')]{S_{12}}'+{\beta_{12 bb}}'
(\tau ') I_{bb}\] 
\[+ {\beta_{12 bg}}'(\tau ') (I_{bg}+I_{dust})\]
where $\beta$ is the usual escape probability.
The cosmic background $I_{bg}$ is
described by a 3 K blackbody and the dust contribution
is given by $I_{dust}$. The emission of
the central star $I_{bb}$ is described by a blackbody spectrum at temperature
$T_{\star}$; a geometrical dilution factor is applied. 
In these formulae, the populations of the {\em 4 levels} are involved. 
If $\tau_{12} < \tau_{34}$, after overlap we have 
${S_{12}}'={S_{34}}'\simeq S_{34}$ and $\tau_{34}' \simeq \tau_{34}$; 
this implies minor changes for $\tau_{34}$ and the $n_{3}$ and 
$n_{4}$ populations. On the other hand, photons from the stronger 
line will be absorbed by the more optically thin line $1-2$, thus 
enhancing the non-equilibrium distribution of the populations in the 
levels 1 and 2.

Such a treatment of the overlaps is oversimplified, but 
allows us to test quickly the effects of this mechanism. 
Our code incorporates three overlaps simultaneously; 
this number can of course be increased. To account for line overlaps, 
the code first derives the populations and opacities separately for 
each isotope without overlap, and then solves again for the radiative transfer, combining 
the overlapping lines. We describe in Sect. 3 how we proceed to 
test the effects of line overlaps.

In fact, non 
local line overlaps have long been recognized as important in the 
excitation of the OH radical, 
and, more recently, in SiO models (cf. Introduction and 
Sect. \ref{discuoverlap}). 
Line overlaps may occur even if two lines are not excited in 
immediately adjacent gas layers; e.g. a Doppler-shifted line emitted from a 
first cloud may interact with another line in a second cloud 
provided that the two clouds have the appropriate relative velocities. 
In a somewhat different process, 
Olofsson \etal (1981, 1985) suggested that the anomalously
weak $v=2, J=2\rightarrow 1$ emission line in stars could 
be explained by non local overlap with the 
H$_{2}$O ${\nu}_{2}~ 12_{75}\rightarrow 
{\nu}_{1}~ 11_{66}$ line. Photons emitted by the 
water line increase the radiation field at the frequency of the 
vibrational transition of SiO creating an excess of absorption for 
this SiO line, and thus destroying the inversion in $v=2, J=2\rightarrow 
1$. This mechanism was recently investigated in detail by Bujarrabal 
\etal (1996).

%
%Figure
%
\begin{figure} [ht]
  \begin{center}
     \epsfxsize=8cm
     \epsfbox{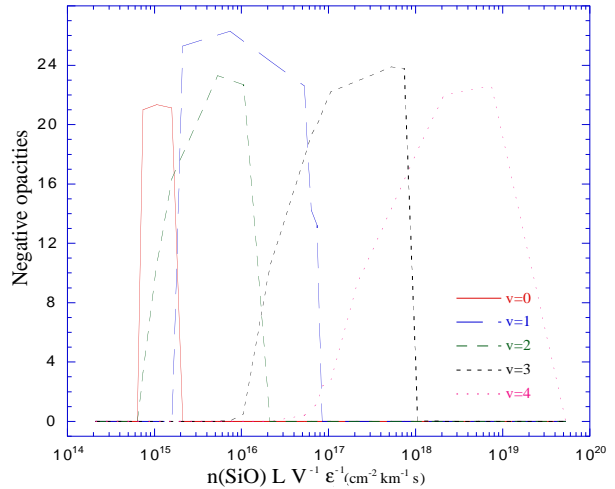}
  \end{center}
  \caption []{\uSiO negative opacities for the transition $J=1\rightarrow
  0$ in $v=0$ to 4. We have taken
  $T_{\star}=2500$ K, $R_{\star}=7.7~10^{13}$ cm,
  the logarithmic velocity gradient $\varepsilon=3$ and
  $\frac{dM}{dt}=5.3~10^{-8}$ M$_{\odot}/$yr. (The $H_{2}$ density
  is fixed to $10^{9}$ cm$^{-3}$ at a distance of $2.8
  R_{\star}$.) $V$ is the expansion velocity.}
  \label{alco10}
\end{figure}

\label{appli}
\subsection{Model validation}
\label{valid}

To test the validity of our code, we have used the same initial
conditions as in Alcolea \etal (1989). (For this comparison, we have not
applied corrections on the collision rates for the limited number of
levels.) The main differences with Alcolea \etal~
are the number of rotational levels
(40 here compared to 12 in Alcolea \etal), the collision rates
(Langer \& Watson 1984 in Alcolea \etal), and the more recent spectroscopic
coefficients used here.
In Figure \ref{alco10}, we plot the $J=1\rightarrow 0$ negative opacities for $v=0$ to 4
versus the ratio $n(SiO) L V^{-1} \varepsilon^{-1}$ (where $L$ is the
Sobolev length, $V$ the expansion velocity, $\varepsilon$ the logarithmic 
velocity gradient). Our results are much similar 
to those of Alcolea \etal ~ and show that the $J=1\rightarrow 0$ 
inversions are well differenciated according to the 
vibrational state. However, some minor discrepancies remain. In 
particular, our work shows that $v=1$ inversions occur 
over a broader range of parameters. We do not obtain increasing 
opacities with increasing $v$; but in Alcolea \etal ~this trend was 
probably the consequence of their smaller number of rotational levels. 
Finally, we note that results in Fig. 2 are similar to those of
Lockett \& Elitzur (1992).

\begin{figure*} [ht]
  \begin{center}
     \epsfxsize=10cm
     \epsfbox{fig3.epsf}
  \end{center}
  \caption []{Negative opacities of newly explained maser lines for
  \dSiO (top) and \tSiO (bottom) with (continuous line) and without
  (dashed line) overlap, versus the distance to the star (one stellar
radius $= 7.7~10^{13}$ cm). The overlaps 
  used in these computations are given in bold face in Table \ref{tablenew2}. 
  The 
  $H_{2}$ density and the abundance $\chi [SiO/H_{2}]$ are 
  respectively $10^{9}$cm$^{-3}$ and 
  $5~10^{-5}$ at a distance of $2.8 R_{\star}$.}
  \label{overfig}
\end{figure*}

%%%%%%%%%%
\scriptsize
\begin{table*} [htb]
 \caption{ \label{tablenew2}Possible explanations of observed 
 maser lines not explained in previous works. The overlaps used in Figs. 
\ref{overfig} and \ref{overCho} are given in bold face. Note the \tSiO $v=1, 
 J=1\rightarrow 0$ maser line newly discovered by Cho \& Ukita (1998).}
 {\small{\begin{tabular}{|l|l|} \hline \hline
{\bf Explained maser lines} & {\bf Overlaps} \\ \hline
\dSiO $v=0, J=5\rightarrow 4$ (weak) & $\bullet$ \dSiO $v=1-0, J=1-0$ + \uSiO 
$v=2-1, J=4-3$ \\ \hline
\dSiO $v=1, J=6\rightarrow 5$ & $\bullet$ \uSiO $v=2-1, J=4-3$ + \dSiO $v=1-0, J=1-0$ \\ 
 & and  \uSiO $v=2-1, J=5-4$ + \dSiO $v=1-0, J=2-1$ \\ 
 & $\bullet$ \uSiO $v=3-2, J=2-3$ + \dSiO $v=2-1, J=5-6$ \\ 
 & $\bullet$ {\bf \dSiO} $\mathbf{v=2-1, J=15-14}$ + {\bf \dSiO} 
 $\mathbf{v=1-0, J=6-5}$ \\ \hline
\dSiO $v=2, J=6\rightarrow 5$ & $\bullet$ {\bf \tSiO} $\mathbf{v=1-0, J=1-0}$ + {\bf \dSiO} 
 $\mathbf{v=1-0, J=3-4}$ \\
 & $\bullet$ \uSiO $v=3-2, J=2-3$ + \dSiO $v=2-1, J=5-6$ \\ \hline
\tSiO $v=0, J=2\rightarrow 1$ & $\bullet$ {\bf \tSiO} 
$\mathbf{v=1-0, J=1-0}$ + {\bf \dSiO} $\mathbf{v=1-0, J=3-4}$ \\
 & $\bullet$  \uSiO $v=2-1, J=3-4$ + \tSiO $v=1-0, J=1-2$\\ \hline
\tSiO $v=0, J=4\rightarrow 3$ & $\bullet$ {\bf \uSiO} $\mathbf{v=2-1, J=3-4}$ + {\bf \tSiO} 
 $\mathbf{v=1-0, J=1-2}$\\ \hline
 \tSiO $v=1, J=1\rightarrow 0$ & $\bullet$ \tSiO $v=1-0, J=1-0$ + 
 \dSiO $v=3-2, J=11-10$ \\ 
  & and \dSiO $v=3-2, J=11-10$ + \uSiO $v=2-1, J=2-3$ \\ 
  & $\bullet$ {\bf \dSiO} $\mathbf{v=1-0, J=3-4}$ + {\bf \tSiO} 
  $\mathbf{v=1-0, J=1-0}$ \\ 
  & $\bullet$ \uSiO $v=2-1, J=3-4$ + \tSiO $v=1-0, J=1-2$ \\ 
\hline \hline
\end{tabular}}}
\end{table*}
\normalsize
%\newpage

\section{Discussion of line overlap effects}
\label{discuoverlap}

Standard radiative and collisional SiO pumping models are able to 
predict several of the observed features of SiO maser emission from 
late-type stars (e.g., Bujarrabal 1994). These models predict similar 
line profiles and smooth line strength variations when one goes up 
the rotational energy ladder. They cannot explain, however, the 
anomalous line intensity ratios observed among several adjacent 
rotational transitions in high $v$ states of \uSiO and in $v=0$ to 3 
states of \dSiO and \tSiO. On the other hand, line overlap effects 
may strongly modify the SiO populations and thus explain apparent 
line anomalies. The most complete work in this domain (Gonz\'alez-Alfonso 
\etal 1996, Gonz\'alez-Alfonso \& Cernicharo 1997) includes a 
non-local treatment of line overlap effects. Their model correctly
predicts several observed features. It also predicts excitation 
of the \dSiO $v=3, J=8\rightarrow 7$ transition which was subsequently 
detected (Gonz\'alez-Alfonso \& Cernicharo 1997) thus demonstrating 
the importance of line overlaps. Nevertheless, excitation of 
several transitions remains unexplained in the work of Gonz\'alez-Alfonso 
\etal~ In particular, emission from \dSiO $v=0, J=5\rightarrow 4$ and 
$v=1,2, J=6\rightarrow 5$ and \tSiO $v=0, J=2\rightarrow 1$ and 
$4\rightarrow 3$ must be understood (see discussion below and Table 
\ref{tablenew2}).

The present model is conceived as a simple tool to quickly investigate 
the impact of various line overlaps between ro-vibrational lines of 
\uSiO, \dSiO and \tSiO. The model incorporates up to 3 simultaneous different 
overlaps, namely six lines, according to the simple prescription 
outlined in Sect. \ref{lesrecouvrements}. Once the maximum line shift allowed 
between two lines is fixed (this depends on $\Delta V_{0}$ defined in 
Sect. \ref{lesrecouvrements}), a specific module of our code searches 
for all possible overlaps. Within the 10 first rotational transitions 
we find a total of 27, 48 and 79 overlaps for $\Delta V_{0}=5, 10$ and 
15 \kms, respectively. There are of course more overlaps involving 
higher J rotational levels, and we have also explored the influence 
of high $J$-level transitions overlapping low $J$-level transitions. 
As briefly explained in Sect. \ref{lesrecouvrements}, the effect of a 
line overlap between two lines strongly depends on the relative 
strength of these two lines. In general, when an overlap occurs 
between \uSiO and \dSiO or \tSiO transitions, the rare isotopic 
transitions which are optically thin tend to be more affected than 
the \uSiO transitions; and the impact on the more optically thin lines 
is higher for lower $v$ states. Our code derives the net optical 
depths compared to the non-overlapping cases.
Taking $\Delta V_{0}=5$ \kms results in 31 overlapping line pairs. These
overlaps affect \uSiO, \dSiO or \tSiO rotational transitions with 
$v \leq 4$ and $J\leq 10$. 
There is however the interesting case of \uSiO $v=2-1, J=3\rightarrow 
4$ which requires $\Delta V_{0} \simeq 10$ \kms to overlap with 
\tSiO $v=1-0, J=1\rightarrow 2$.

Because of radiative pumping through high $v$ levels, we
stress that very different ($v,J$) levels are connected. That is why 
some infrared overlaps have sometimes a strong
influence on ro-vibrational levels not involved in these overlaps: because of 
this excitation mechanism numerous very different levels are 
connected together.

We have grouped our results into four categories: ({\em i}) the overlaps 
invoked by Gonz\'alez-Alfonso \etal (1996) and 
Gonz\'alez-Alfonso \& Cernicharo (1997) to explain 
some important newly discovered rotational 
transitions; ({\em ii}) the overlaps which may account for the behaviour of 
maser rotational transitions explained elsewhere by other models and 
other overlaps; 
({\em iii}) the overlaps providing explanations for the maser lines not yet 
explained; and ({\em iv}) the overlaps suggesting a search for new maser 
lines. 

$\bullet$ In the first category it is worth mentioning that the overlap 
of \uSiO $v=2-1, J=20-21$ with \dSiO $v=3-2, J=8-9$, although it 
involves relatively high $v$ states for \dSiO, results in strong 
enhancement of the \dSiO $v=3, J=8-7$ transition. This was also 
predicted by Gonz\'alez-Alfonso \etal (1996). We have also verified 
that two line overlaps, \uSiO $v=2-1, J=2-3$ with \dSiO $v=3-2, 
J=11-10$, and \dSiO $v=3-2, J=11-10$ with \tSiO $v=1-0, 
J=0-1$, result in a clear enhancement of the \dSiO $v=3, 
J=11\rightarrow 10$ line intensity. Another example is 
the enhancement of the \tSiO $v=2, J=8\rightarrow 7$ transition
resulting, as in Gonz\'alez-Alfonso \etal (1996), from the effects of 3
simultaneous line overlaps. 
%%%%%%%%%%
%\newpage
\scriptsize
\begin{table} [ht]
\caption{ \label{tabledetect}Predicted new maser transitions. The lines 
in bold face have not been searched for maser emission as far as we are aware.}
 {\small{\begin{tabular}{ccc} \hline \hline
{\bf \uSiO} & $v=4$ & $\mathbf {J=3\rightarrow 2}$ \\ 
 & & $J=4\rightarrow 3$ \\ \hline
{\bf \dSiO} & $v=0$ & $J=4\rightarrow 3$ \\ 
 & $v=1$ & $J=2\rightarrow 1$ \\ 
 & & $J=5\rightarrow 4$ \\
 & {\bf $v=2$} & $\mathbf {J=1\rightarrow 0}$ \\
 & & $\mathbf {J=3\rightarrow 2}$ \\
 & {\bf $v=3$} & $\mathbf {J=2\rightarrow 1}$ \\
 & & $\mathbf {J=4\rightarrow 3}$ \\ \hline
{\bf \tSiO} & $v=1$ & $J=2\rightarrow 1$ \\
 & {\bf $v=2$} & $\mathbf {J=2\rightarrow 1}$ \\
\hline \hline
\end{tabular}}}
\end{table}
\normalsize
%\newpage

$\bullet$ We have successfully explained (case ({\em ii})) the 
enhancement of the \uSiO $v=3, J=1\rightarrow 0$ and $v=4, 
J=5\rightarrow 4$ transitions, and of the \tSiO $v=0, J=5\rightarrow 4$ 
transition. Standard modelisation does not easily invert the \uSiO $v=3, 
J=1\rightarrow 0$ line although it is observed in a variety of 
sources (Scalise \& L\'epine 1978, Alcolea \etal 1989). One overlap 
may lead to the excitation of this line: 
\uSiO $v=3-2, J=2-3$ with \dSiO $v=2-1, 
J=5-6$. On the other hand, observations indicate that the \uSiO $v=3, 
J=2-1$ line should be quenched. This could result from four different 
overlaps including the low $J$ rotation level overlap of \uSiO 
$v=2-1, J=3-4$ with \tSiO $v=1-0, J=1-2$. This overlap is also 
consistent with the extinction of the $v=3, J=2-1$ line and with the 
enhancement of the $v=4, J=5\rightarrow 4$ line as observed by 
Cernicharo \etal (1993) in VY CMa. 
Our model easily leads to maser emission of the $v=0,
J=1\rightarrow 0$ transition. This is observed in several stars for \dSiO
(Alcolea \& Bujarrabal 1992) and \tSiO (Cernicharo \& Bujarrabal 1992).

\begin{figure} [ht]
  \begin{center}
     \epsfxsize=7cm
     \epsfbox{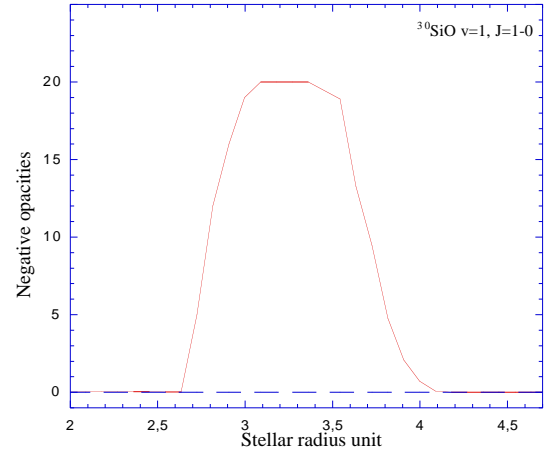}
  \end{center}
  \caption []{Negative opacities for the \tSiO $v=1, J=1\rightarrow 0$ maser line
   with (continuous line) and without
  (dashed line) overlap, versus the distance to the star (one stellar
  radius $= 7.7~10^{13}$ cm). The overlap 
  used in this computation is given in bold face in Table \ref{tablenew2}. The 
  $H_{2}$ density and the abundance $\chi [SiO/H_{2}]$ are 
  respectively $10^{9}$cm$^{-3}$ and 
  $5~10^{-5}$ at a distance of $2.8 R_{\star}$.}
  \label{overCho}
\end{figure}

$\bullet$ Table \ref{tablenew2} (case ({\em iii})) 
and Fig.\ref{overfig} suggest explanations of the \dSiO and \tSiO 
maser transitions not explained in the work of Gonz\'alez-Alfonso 
\etal (1996). Only
overlaps can explain inversions of the \tSiO $v=0, J=2\rightarrow 1,
4\rightarrow 3$ (not explained in previous works) or $J=5\rightarrow 4$ transitions. Line overlaps are
also required to explain several \dSiO $v=1$ and 2 rotational
transitions. For the $v=1, J=6\rightarrow 5$ line observed by
Cernicharo \& Bujarrabal (1992), Table \ref{tablenew2} suggests that
it may be explained by several overlaps. Of special interest is the
simultaneous overlap of \uSiO $v=2-1, J=4-3$ with \dSiO $v=1-0,
J=1-0$ and of \uSiO $v=2-1, J=5-4$ with \dSiO $v=1-0, J=2-1$. The same
arguments apply to $v=2, J=6\rightarrow 5$ which has also been
observed by Cernicharo \& Bujarrabal. The overlap of \dSiO $v=1-0,
J=3-4$ with \tSiO $v=1-0, J=1-0$ is clearly important because it
enhances simultaneously the observed $J=4-3$ and $2-1$ transitions. 
It is interesting to note that we have 
succeeded in producing weak maser emission in \dSiO 
$v=0, J=5\rightarrow 4$ 
(never explained before). Another important result concerns the 
\tSiO $v=1, J=1\rightarrow 0$ maser emission produced by our model 
with different overlaps (see Table \ref{tablenew2} and Fig. \ref{overCho}); 
this line was recently discovered by Cho \& Ukita (1998) in TX Cam.

$\bullet$ Table \ref{tabledetect} lists some relatively strong maser transitions
as predicted by our model (case ({\em iv})). 
Note that the \dSiO and \tSiO
$v=1, J=2\rightarrow 1$ maser lines have also been predicted by 
Rausch \etal (1996) on the basis of an optical pumping model. 
Four transitions
(\uSiO $v=4, J=3\rightarrow 2$, \dSiO $v=1, J=2\rightarrow 1$ and
$v=3, J=4\rightarrow 3$, \tSiO $v=2, J=2\rightarrow 1$)
are easily excited even without line
overlaps; however, all other transitions require line overlaps for 
being detectable. For \dSiO $v=1$ we find that 6 different overlaps of 
ro-vibrational lines of \uSiO with \dSiO tend to enhance the 
$J=5\rightarrow 4$ transition.
Six transitions (given in bold face in Table \ref{tabledetect}:
\uSiO $v=4, J=3\rightarrow 2$, \dSiO $v=2, 
J=1\rightarrow 0, 3\rightarrow 2$ and $v=3, J=2\rightarrow 1, 4\rightarrow 
3$, and \tSiO $v=2, J=2\rightarrow 1$) have not yet been searched 
for maser emission as far as we are aware. However, all other transitions 
in Table 2 have already been observed without success by 
several authors (e.g., Cernicharo \& Bujarrabal 1992).
Nevertheless, further
investigations of these lines would be useful, especially as
\dSiO $v=1, J=2 \rightarrow 1$ was recently detected in one star 
(J.R. Pardo, private communication).

Finally, we stress that other types of line
overlaps may simultaneously play a role in the anomalous excitation 
of \uSiO, \dSiO and \tSiO. The infrared lines of water are especially 
important in this context because water is an abundant species in the 
late-type stars where SiO is present. Apart from the 
H$_{2}$O-SiO line overlap discussed by Olofsson \etal (1981, 1985) and 
Bujarrabal \etal (1996), there are several other infrared 
transitions of ortho- or para-H$_{2}$O close to ro-vibrational 
transitions of SiO and isotopes which would deserve modelisation. 

\section{Conclusion}

We have presented results for a new and simple model of SiO masers which incorporates
both radiative and collisional pumpings as well as line overlap effects.
Radiative pumping includes stellar radiation and circumstellar grains
although their extinction coefficient is not well known. Collisional pumping
involves molecular hydrogen as well as atomic hydrogen (under the form of simply
guessed H-SiO collision rates). Our model involves 40 rotational levels
for each of the first five vibrational states and derives the
populations and opacities for the three species \uSiO, \dSiO and \tSiO, and for a
variety of circumstellar conditions. 
Several low and high $J$ rotational transitions are easily inverted
in the vibrational states $v=1,2,3$ and 4 as observed in many late-type stars. 
Prediction of relative line intensities must account for line overlap effects 
which have been shown to play an important role in the excitation of SiO
masers.

Our model incorporates in a very simple way the effects of line overlaps between
ro-vibrational lines of \uSiO, \dSiO and \tSiO. Three different overlaps may be
treated simultaneously, and we have examined the impact on the level populations
and opacities of 31 overlapping line pairs. We have found that line overlaps are
important to enhance \uSiO $v=3, J=1 \rightarrow 0$ and $v=4, J=5 \rightarrow 4$
maser emission or to turn off $v=3, J=2 \rightarrow 1$ line emission. Various overlaps
may lead to weak \dSiO $v=0, J=5 \rightarrow 4$ line emission and strong
\dSiO $v=1$ and 2, $J=6 \rightarrow 5$ emission; these lines had not been explained
in other works. More generally, several \dSiO $v=1$ and 2 maser lines are well explained
with line overlaps. We have also suggested possible explanations
to several \tSiO maser lines including the newly discovered transition
$v=1, J=1 \rightarrow 0$. In addition, we have predicted that several new maser transitions
for \uSiO ($v=4$), \dSiO ($v=0$ to 3) and \tSiO ($v=1$ and 2) could result
from line overlap effects. 

\acknowledgements{We thank the referee, Dr. W.H. Kegel, for useful comments 
on the manuscript.}

\normalsize
\end{document}